\documentclass{emulateapj}
\pdfoutput=1
\usepackage{epsfig,natbib}
\usepackage{amsmath,amsfonts,amssymb}
\usepackage{mathptmx}
\usepackage{apjfonts}
\newcounter{species}

\begin{document}

\shorttitle{Probing MHD Instabilities with Faraday Rotation}
\title{Using Faraday Rotation to Probe MHD Instabilities in Intracluster Media}

\author{Tamara Bogdanovi\'c\altaffilmark{1,2}, Christopher
S. Reynolds\altaffilmark{1}, and Richard Massey\altaffilmark{3}}

\altaffiltext{1}{Department of Astronomy, University of Maryland,
College Park, MD 20742-2421, e-mail: {\tt tamarab, chris@astro.umd.edu}}

\altaffiltext{2}{Einstein Postdoctoral Fellow}

\altaffiltext{3}{Royal Observatory Edinburgh, Blackford Hill,
Edinburgh EH9 3HJ, UK}

\begin{abstract}

It has recently been suggested that conduction-driven
magnetohydrodynamic (MHD) instabilities may operate at all radii
within an intracluster medium (ICM), and profoundly affect the
structure of a cluster's magnetic field.  Where MHD instabilities
dominate the dynamics of an ICM, they will re-orient magnetic field
lines {\it perpendicular} to the temperature gradient inside a cooling
core, or {\it parallel} to the temperature gradient outside it.  This
characteristic structure of magnetic field could be probed by
measurements of polarized radio emission from background sources.
Motivated by this possibility we have constructed 3-d models of a
magnetized cooling core cluster and calculated Faraday rotation
measure (RM) maps in the plane of the sky under realistic observing
conditions.  We compare a scenario in which magnetic field geometry is
characterized by conduction driven MHD instabilities to that where it
is determined by isotropic turbulent motions. We find that future
high-sensitivity spectro-polarimetric measurements of RM, such as will
be enabled by the {\it Expanded Very large Array} and {\it Square
Kilometer Array} can distinguish between these two cases with
plausible exposure times. Such observations will test the existence of
conduction-driven MHD instabilities in dynamically relaxed cooling
core clusters. More generally, our findings imply that observations of
Faraday RM should be able to discern physical mechanisms that result
in qualitatively different magnetic field topologies, without a priori
knowledge about the nature of the processes.
\end{abstract}


\section{Introduction}

Understanding the role and evolution of magnetic fields in clusters of
galaxies is of significant importance for many questions including the
origin of cluster magnetic fields, the interaction of active galactic
nuclei (AGN) with the intracluster medium (ICM), and physical
processes operating within the ICM plasma.

One of the key techniques used to obtain information about the
strength and structure of cluster magnetic fields is the analysis of
Faraday rotation from polarized radio sources located behind and
within clusters. The Faraday effect rotates the polarization plane of
electromagnetic waves traveling through a magnetized plasma, by an
amount proportional to the (square of the) wavelength, plasma density,
and the strength of the field in the direction of propagation.  The
sources' intrinsic polarization need not be known, as the effect can
be observed as a characteristic wavelength-dependent rotation measure
(RM) signature. Observations of a few nearby clusters have established
the presence of magnetic fields with typical strengths of ${\rm
few\,\mu G}$ in non-cool core clusters and in excess of 10${\rm\, \mu
G}$ in the centers of cool core clusters \citep{ct02,gf04,fg08}.
Detailed high resolution RM images of radio galaxies in merging and
cooling core clusters indicate that the RM distribution is
characterized by patchy structures of a few kpc in size. Furthermore,
the RM distribution appears in general in good agreement with the RM
expected for magnetic fields geometry characterized by turbulent
motions with a power-law power spectrum \citep[][for
e.g.]{ev03b,murgia04, guidetti08,laing08,bonafede10}. This base of
knowledge will be greatly enhanced by radio observatories such as the
{\it Expanded Very Large Array}\footnote{http://science.nrao.edu/evla}
({\it EVLA}) and {\it Square Kilometer
Array}\footnote{http://www.skatelescope.org} ({\it SKA}), which will
provide the sensitivity to study both background and embedded sources
with sufficient density to map out the RM patterns in many and more
distant clusters. This will allow to study the structure of magnetic
field in ``quiescent'' regions of clusters, away from radio galaxies,
as well as the effect of the radio-mode feedback in their vicinity.

The interpretation of these results can strongly benefit from advances
in the theoretical understanding of the dynamics of weakly magnetized,
low density plasmas. The ICM is a dilute plasma, i.e., the
gyro-frequency of both the electrons and the ions is much greater than
the collision frequency. The principal effect of this is highly
anisotropic thermal conduction that fundamentally alters the classical
Schwarzschild criterion for convection \citep{balbus00}. Under these
circumstances, the outer regions of clusters where temperature
decreases with radius may become susceptible to a conduction-driven
{\it magneto-thermal instability} (MTI). The properties of this
instability have been studied in MHD simulations
\citep{ps05,ps07,psl08}. \citet{psl08} found that MTI can profoundly
affect the temperature distribution in the outer regions of a cluster
in only a few billion years, by re-orienting the lines of magnetic
field to be preferentially radial, resulting in a thermal conduction
at a high fraction of the Spitzer conductivity.

The temperature profile in the centers of cooling-core galaxy clusters
makes them stable to MTI. However, \citet{quataert08} found that
cooling cores are characterized by a sister {\it heat-flux buoyancy
instability} (HBI) that arises in regions where temperature increases
with radius. An initial investigation of nonlinear evolution of the
HBI \citep{pq08} indicates that its primary effect is to re-orient the
lines of magnetic field perpendicular to the temperature gradient, and
thus strongly suppress heat conduction. The properties of HBI have
been investigated in global 3-d models of isolated cooling core
clusters \citep{pqs09,bogdanovic09}. They suggest that, once the
magnetic field lines have been wrapped into spherical surfaces
surrounding the core, the effective thermal conduction is suppressed
to a small fraction of the Spitzer value, leading to insulation of the
core from further conductive heating, and to a subsequent thermal
collapse.

Very recently, \citet{br10} discovered an associated pair of {\it
overstabilities} that affect precisely those configurations that are
stable to the well-established HBI and MTI. They predict that
configurations which tend to result from the non-linear evolution of
the HBI have $g$-modes that are driven overstable by radiative
loses. On the other hand, configurations which tend to result from the
non-linear evolution of the MTI have $g$-modes that are driven
overstable by the conductive heat flux. The effects of these
overstabilities for the ICM plasma thermodynamics and the properties
of magnetic field are yet to be understood. We do not consider them in
this work and instead focus on MTI and HBI instabilities.

The combination of MTI and HBI instabilities should lead to a
characteristic structure of magnetic fields in some cooling core
clusters.  If these conduction-driven instabilities dominate the
dynamics of the ICM, the magnetic field lines would be preferentially
oriented radially in the outer region and azimuthally within the
cooling core. This may not be the case in all clusters -- it has
recently been shown that MHD instabilities can be overwhelmed by even
moderate levels of driven turbulence \citep{ro09,pqs10}, such as might
result from sub-cluster mergers, motions of cluster member galaxies,
or various forms of AGN feedback. Which (if any) physical mechanism
dominates in the ICM depends sensitively upon the magnitude and
distribution of turbulence, which is currently only poorly understood.
Motivated by this question and guided by the MHD simulations, we have
constructed 3-d models of clusters for two distinct cases and
simulated the RM maps that might be observed with high sensitivity
radio polarization measurements.

This paper is organized as follows. We describe the ingredients of our
cluster models in \S~\ref{S_model} and the main properties of derived
RM maps in \S~\ref{S_RMmaps}. We discuss the importance of
depolarization and diffuse emission in \S~\ref{S_depolarization}, then
present the discussion with conclusions in \S~\ref{S_discussion}.

\section{Cluster models}\label{S_model}

\begin{deluxetable*}{ccccccccc} 
\tablecaption{Summary of model properties\label{table}}
\tablewidth{0pt}
\tablecolumns{13}
\tablehead{
\colhead{Model} &
\colhead{Observatory} &
\colhead{Scenario} &
\colhead{$N_s$} & 
\colhead{$d_s$} & 
\colhead{\underline{$\sum |a_{n,0}|$}} &
\colhead{\underline{$\sum |a_{n,1}|$}} &
\colhead{\underline{$\sum |a_{n,2}|$}} &
\colhead{\underline{$\sum |a_{n,3}|$}} \\
\colhead{} &
\colhead{} &
\colhead{} &
\colhead{} & 
\colhead{(kpc)} &
\colhead{$\sum |a_{n,m}|$} &
\colhead{$\sum |a_{n,m}|$} &
\colhead{$\sum |a_{n,m}|$} &
\colhead{$\sum |a_{n,m}|$} 
}
\startdata
  & SKA  &  1h, low    &  449  & 28  &  8.1\% & 39.1\% & 15.3\% & 2.1\% \\
  & SKA  &  1h, high   &  7117 & 7   & 17.3\% & 45.1\% & 15.4\% & 1.9\% \\ 
A & SKA  &  100h, low  &  2250 & 13  &  7.8\% & 43.5\% & 17.6\% & 2.5\% \\
  & SKA  &  100h, high & 89592 & 2   & 14.3\% & 50.0\% & 17.7\% & 0.7\% \\
  & EVLA &  9h, low    &  65   & 75  & 11.1\% & 28.2\% & 10.5\% & 3.9\% \\
  & EVLA &  9h, high   &  337  & 33  &  9.0\% & 39.5\% & 15.8\% & 2.4\% \\
\hline\\
   & SKA  &  1h, low    &  449  & 28 & 25.3\% & 15.6\% & 11.6\% & 5.4\% \\
   & SKA  &  1h, high   &  7117 & 7  & 11.9\% & 12.1\% &  9.4\% & 5.6\% \\ 
B  & SKA  &  100h, low  &  2250 & 13 & 19.3\% & 16.0\% & 10.4\% & 5.7\% \\
   & SKA  &  100h, high & 89592 & 2  &  9.2\% &  7.5\% &  9.1\% & 5.7\% \\
   & EVLA &  9h, low    &  65   & 75 & 31.1\% & 16.6\% & 13.2\% & 4.4\% \\
   & EVLA &  9h, high   &  337  & 33 & 20.5\% & 19.6\% & 11.0\% & 7.6\% 
\enddata
\footnote[]{Model -- A (B) corresponds to the instability-dominated
(turbulence-dominated) cluster model; Observatory -- one of the two
radio-observatories considered in the paper; Scenario -- see the text;
$N_s$ -- number of background polarized sources within
600~kpc$\times$600~kpc area; $d_s$ -- mean separation of polarized
background sources; ratios -- fraction of the total RM intensity in
circularly symmetric shapelet multipoles.}
\end{deluxetable*}

\subsection{Intracluster medium and magnetic fields}\label{S_cluster_model}

Any magnetized cluster acts as a Faraday screen for polarized sources
located behind a cluster or in the cluster itself. Modeling the RM
signature of such a cluster will require several main ingredients: the
density and temperature distribution of electrons in the ICM of the
cluster, the 3-d structure of the magnetic field, and the density and
fluxes of polarized background sources. We construct models in a
Cartesian coordinate system $(x,y,z)$ with a cubic spatial domain
defined by $x=\pm L$, $y=\pm L$, $z=\pm L$, where $L=300$~kpc. For the
electron density and temperature distributions we adopt analytic
approximations based on the XMM-Newton observations of the Perseus
cluster that capture the radial behavior of these two parameters
\citep[we use expressions from][and scale them to our
assumed cosmology]{churazov03}.
\begin{eqnarray}
n_e & = & \frac{3.9\times10^{-2}}{\left[1+(r/r_a)^2\right]^{1.8}} +
\frac{4.05\times10^{-3}}{\left[1+(r/r_b)^2\right]^{0.87}}\; {\rm cm^{-3}} \;,\label{eq_ner}\\
T_e & = & 7\frac{\left[1+(r/r_c)^3\right]}{\left[2.3+(r/r_c)^3\right]} {\rm keV}\;,\label{eq_Ter}
\end{eqnarray} 
where $r = (x^2+y^2+z^2)^{1/2}$ is in units of kiloparsecs and $r_a =
56.5$~kpc, $r_b = 197.7$~kpc, and $r_c = 70.6$~kpc. The central electron
number density and temperature are $n_e(0)=4.3\times10^{-2}\,{\rm
cm^{-3}}$ and $T_e(0)=3\,{\rm keV}$, respectively.

We consider two models of the cluster's magnetic field. Our first
model (model A) is motivated by the results of recent MHD simulations
of instabilities in the ICM, where the magnetic field is
preferentially azimuthal within the cooling core and radial outside of
this region. Hence, in a spherical polar coordinate system
$(r,\theta,\phi)$, where $\theta=0$ is aligned along the $z$-axis, the
field structure within the cooling core, $r\leq r_c=200$~kpc, is
described in terms of only $\theta$ and $\phi$ components
\citep{bogdanovic09}:
\begin{eqnarray}
B_{\theta} & = & 2B_0 (1 + \sin(2\pi r/r_1)) \sin\theta \cos(2\phi), \label{eq_Btheta} \\ 
B_{\phi} & = & 2B_0 (1 + \sin(2\pi r/r_2)) \sin(3\theta) - \label{eq_Bphi} \\
         &   & B_0 (1 + \sin(2\pi r/r_1)) \sin(2\phi) \sin(2\theta), \nonumber
\end{eqnarray} 
$B_0(r)$ is chosen so that the value of the plasma parameter, $\beta =
8\pi n_e\, k\, T_e /B_0^2 = 100$, is constant everywhere within the
cooling core region. This implies $B_0(0) = 7.3\,{\rm \mu G}$ at the
cluster center, the field strength in the range $2.3\times
10^{-5}-36.2\,{\rm \mu G}$, and mean strength over the computational
volume of $3.3\,{\rm \mu G}$. $r_1 = 7.5$~kpc and $r_2 = 24$~kpc are
the coherence lengths defining characteristic radial scales on which
magnetic field vector changes direction. The magnetic field thus
changes direction 8 to 26 times across the cool core radius. In
addition to these two scales, equations~\ref{eq_Btheta} and
\ref{eq_Bphi} also capture the field geometry expected to arise as a
consequence of the HBI: field lines wrapped onto the spherical
surfaces within the cool core. This implies that in model A magnetic
field reversal also occurs on a range of spatial scales associated
with the spheres of different radii and up to the size of the cool
core, $2 r_c = 400$~kpc. It follows that 400~kpc is the maximum scale
for magnetic field fluctuations in model A.

The magnetic field structure outside of the cooling core, $r>r_c$, is
described by a radial component,
\begin{eqnarray} \label{eq_Br}
B_r = \frac{B_{r0}}{r^2}, 
\end{eqnarray}
where $B_{r0}$ is defined from the condition $B^2_{r}(r_c) = \langle
B^2_\theta(r_c)+ B^2_\phi(r_c) \rangle$. The simple analytic form for
the field structure given by equations
\eqref{eq_Btheta}--\eqref{eq_Br} satisfies the condition $div\,{\bf
B}=0$ everywhere except at $r=r_c$, where the field lines are
transitioning from azimuthal to radial. In a realistic case, this
transition would be more gradual and also divergence free. We
nevertheless expect that, except around $r\approx r_c$, our model
should capture the salient properties of the magnetic field of a
cluster affected by MHD instabilities across a wide range of radii.

In our second model (model B), magnetic field lines are randomly
tangled reflecting a different physical scenario in which the field
geometry is set by the action of isotropic turbulence. While this
theoretical hypothesis is most likely idealized, it is physically
motivated and well rooted in theoretical practice, so we use it as a
control case to model A.  Following \citet{roet99}, we initialize the
field geometry by defining a magnetic field potential in Fourier space
of $\tilde{A}(k)=\tilde{A}_0\,k^{-\alpha}$, where each Cartesian
factor $\tilde{A}_0$ has an amplitude drawn from a Gaussian
distribution and a random phase, assuring uncorrelated modes. We
convert this into real space via a three-dimensional fast Fourier
transform (FFT). The tangled magnetic field is then calculated as
${\bf B}=\nabla\times {\bf A}$.  We adopted $\alpha = 17/6$, which
results in the Kolmogorov-like power spectrum $B^2\propto
k^{2(1-\alpha)} \propto k^{-11/3}$. The smallest and largest magnetic
structures produced by this power spectrum have the scales of
$\lambda_{\rm min} = 2\pi/k_{\rm max}= 7.5$~kpc and $\lambda_{\rm max}
= 2\pi/k_{\rm min}= 600$~kpc, respectively. The magnitude of magnetic
field is normalized in such way that the azimuthally averaged magnetic
energy comprises 1\% of the thermal energy of the gas at all radii
(i.e., $\beta =100$, same as in model A). This implies the mean
magnetic field strength decreasing with the distance from the cluster
center and $B_0(0) = 7.3\,{\rm \mu G}$.  The field strength in model~B
varies between $8.3\times10^{-8}-29.8\,{\rm \mu G}$ and its mean
amplitude over the computational volume is $1.3\,{\rm \mu G}$.

\subsection{Observational scenarios}

We place the model cluster at the redshift of the Perseus cluster,
which is a suitable prototype for a nearby cooling core cluster. At
the distance of NGC~1275 \citep[$z=0.0176$;][]{strauss92}, the central
bright galaxy in Perseus, $1\arcsec$ corresponds to
$353\,$pc.\footnote{Throughout this paper we adopted a $\Lambda$CDM
cosmology with $H_0 = 71\, {\rm km\,s^{-1}\,Mpc^{-1}}$, $\Omega_m =
0.27$, and $\Omega_\Lambda = 0.73$.} To simulate future, high
sensitivity RM surveys, we adopt the planned capabilities of the soon
to be fully operational {\it EVLA} and the next-generation radio
interferometer, {\it SKA}, at 1.4~GHz.

{\it EVLA} consists of 27 25-meter diameter antennas that will provide
an order of magnitude increase in sensitivity above the existing {\it
Very Large Array} after it is upgraded with new receivers and
electronics. Together with improved resolution and imaging, these new
capabilities make {\it EVLA} an important tool for studies of Faraday
rotation in the near future.\footnote{Currently, {\it EVLA} is on
schedule for completion at the end of 2012.} Its technical
specifications include a continuous coverage with full polarization
capabilities between frequencies of 1 and 50~GHz, a field of view
(FOV) of $\sim ~0.25$~sq. deg at 1.4~GHz, and angular resolution as
high as $\theta_{\rm EVLA}= 1.3\arcsec$ (achievable in the largest
array configuration; A-configuration) also at 1.4~GHz ($\lambda =
21$~cm). Furthermore, observations of the continuum emission with the
{\it EVLA} are expected to achieve the r.m.s. noise limit of $1.6\,
{\rm \mu Jy/beam}$ between 1 and 2~GHz in about 9h of exposure. In the
next step we will use this sensitivity limit to estimate the density
of the background polarized sources on the sky that will be seen by
the {\it EVLA} at 1.4~GHz.\footnote{Information about the {\it EVLA}
capabilities and specifications was obtained from
http://www.vla.nrao.edu/astro/guides/vlas/current/ .}

With further improvements in sensitivity and survey speed over current
instruments, {\it SKA} will be ideally equipped to study the origin
and evolution of cosmic magnetism in the future
\citep{cr04,dewdney09,krause09}.\footnote{At this point a commencement
of full science operations with the {\it SKA} is planned for 2020.}
Planned technical specifications for the {\it SKA} include a square
kilometer collecting area, continuous frequency coverage from 70~MHz
to 25~GHz, a FOV of 1~sq. deg at 1.4~GHz, and angular
resolution better than $\theta_{\rm SKA}=1\arcsec$ at the same
frequency \citep{sch07,taylor08}.  Note however that the frequency
range for RM studies with the {\it SKA} will be relatively wide ($\sim
0.3-10$~GHz), and that the exact FOV and angular resolution depend on
the frequency. Observations with the {\it SKA} are expected to achieve
an r.m.s. noise of $0.1\, {\rm \mu Jy/beam}$~area and $0.01\, {\rm \mu
Jy/beam}$~area at 1.4~GHz within 1h and 100h of integration,
respectively \citep{cr04}.

\begin{figure*}[t]
\center{
\includegraphics[width=0.75\textwidth]{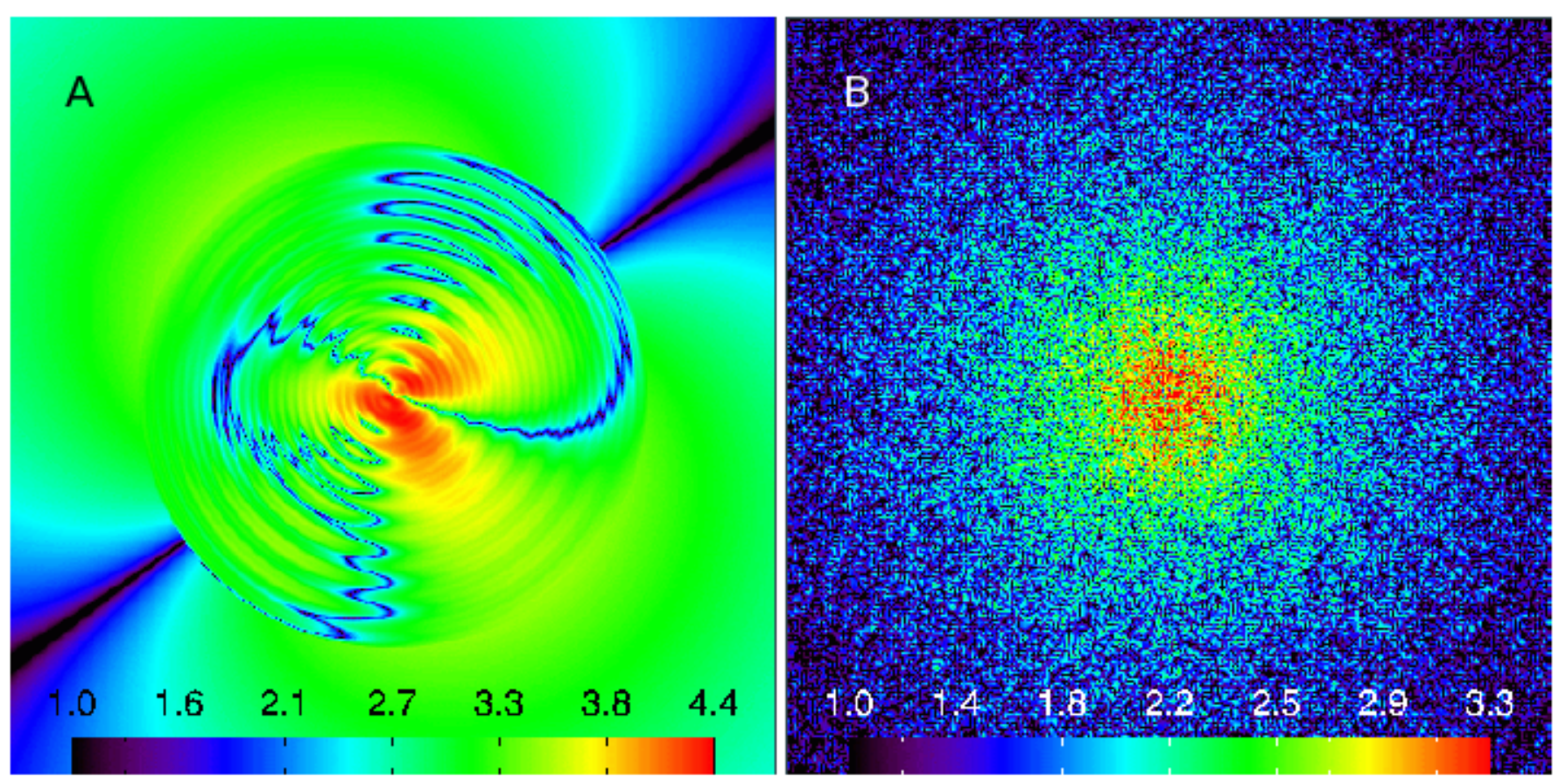} 
\caption{Simulated {\it continuous} Faraday rotation measure maps for
an instability-dominated (model A) and turbulence-dominated (model B)
cluster. Panel size is 600~kpc on a side and the color scale shows
$\log\left|{\rm RM}\right|$.  In model A, two distinct regions are
visible: the magnetic field is dominated by HBI inside the cooling
core, and MTI outside.  The diagonal feature at large radii is an
artifact of the model. In model B, the cluster's magnetic field is
randomly tangled, producing a patchy RM
distribution.}\label{fig_mapsAB}}
\end{figure*}
\begin{figure*}[t]
\center{
\includegraphics[width=0.75\textwidth]{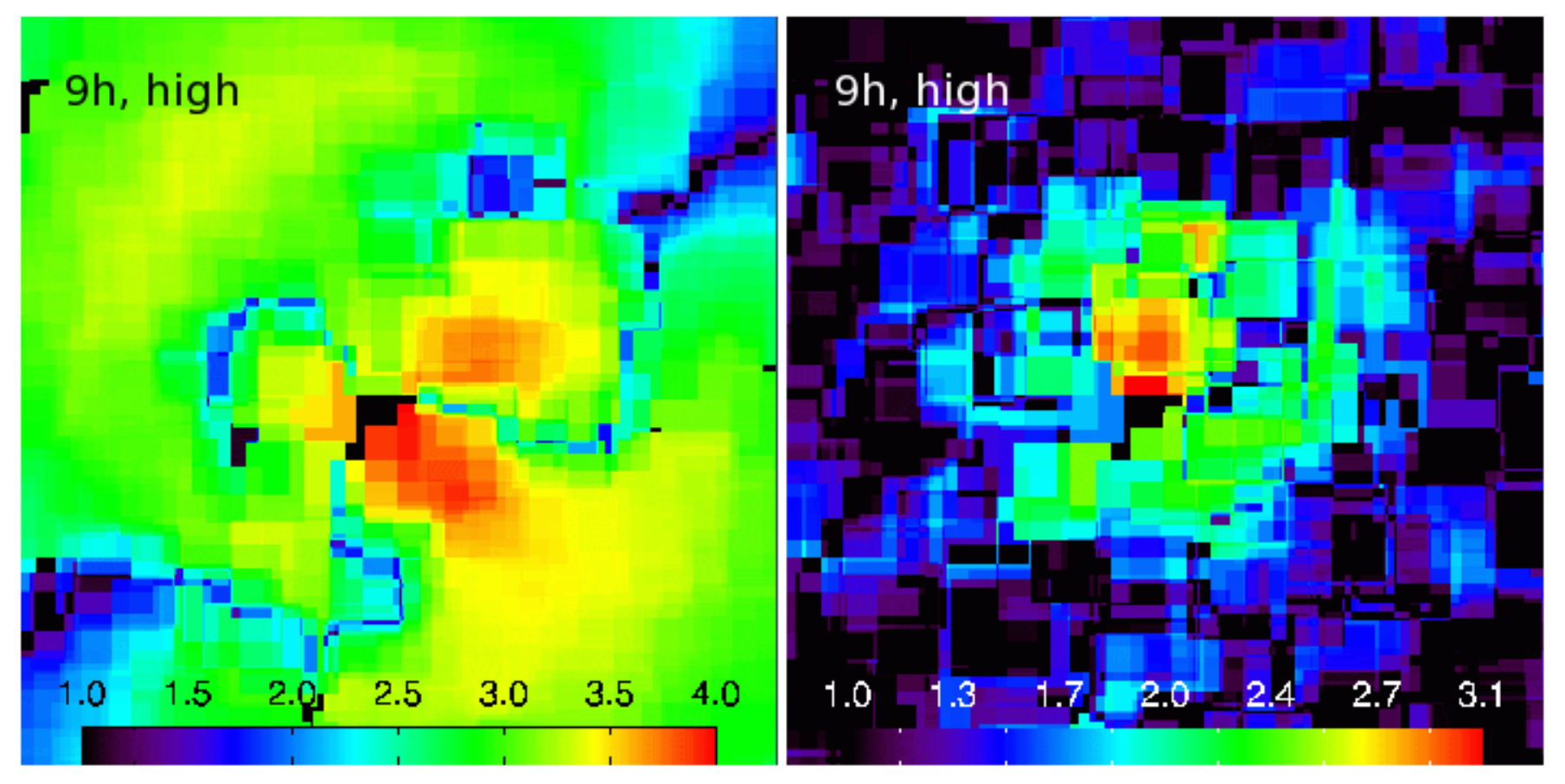} 
\caption{Faraday rotation measure maps simulated for a 9 hour exposure
with {\it EVLA } in a high source count observational scenario. Panels
show instability-dominated cluster considered in model A (left) and
turbulence-dominated cluster from model B (right). Panels show the
same region and color scaling as in
Figure~\ref{fig_mapsAB}.}\label{fig_evla}}
\end{figure*}
\begin{figure*}[t]
\center{
\includegraphics[width=0.75\textwidth]{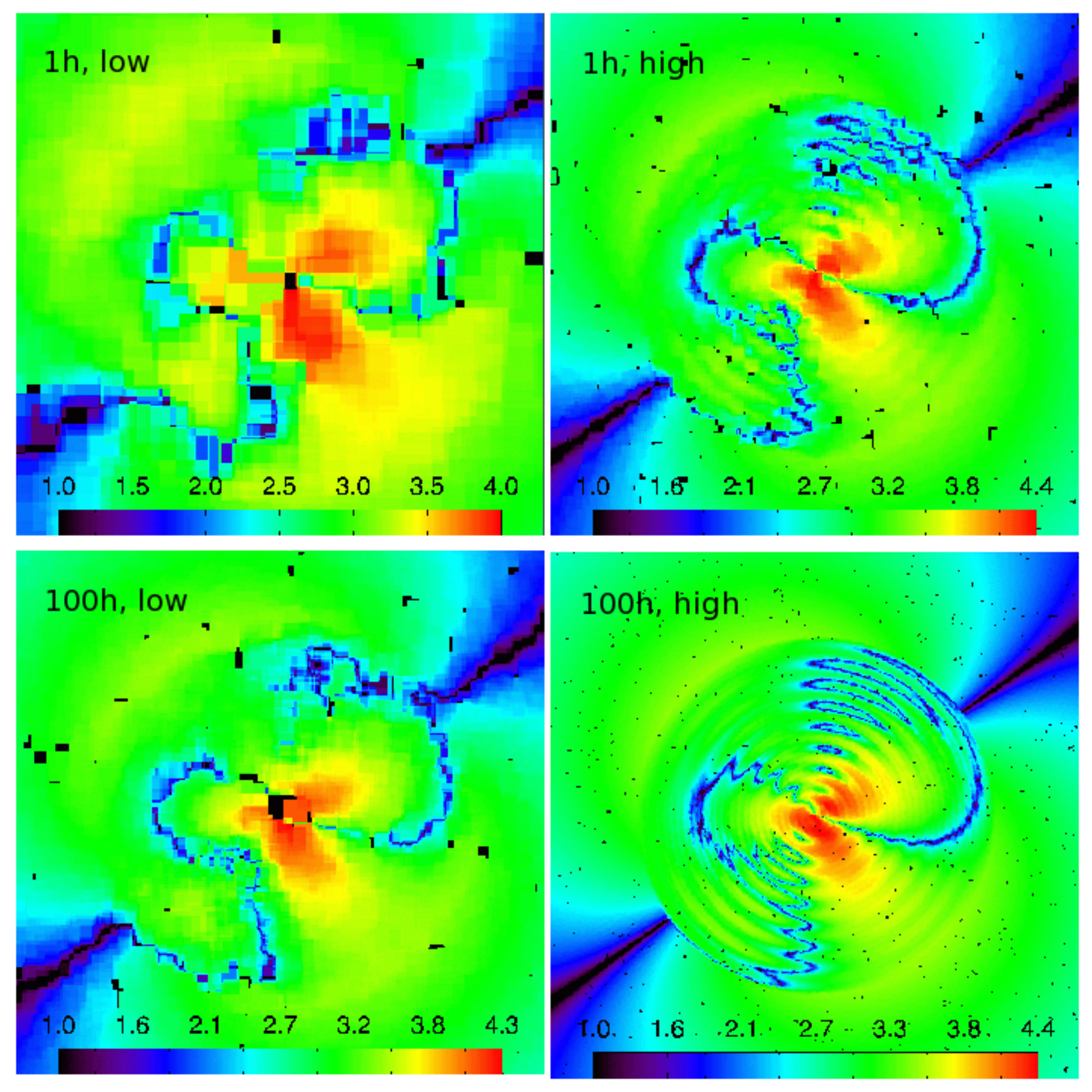} 
\caption{ Simulated {\it SKA} Faraday rotation measure maps for an
instability-dominated (model A) cluster. The four observing scenarios
include short or long exposure times, with low or high background
source densities. Panel size and color scale are the same as in
previous figures.}\label{fig_hbi}}
\end{figure*}
\begin{figure*}[t]
\center{
\includegraphics[width=0.75\textwidth]{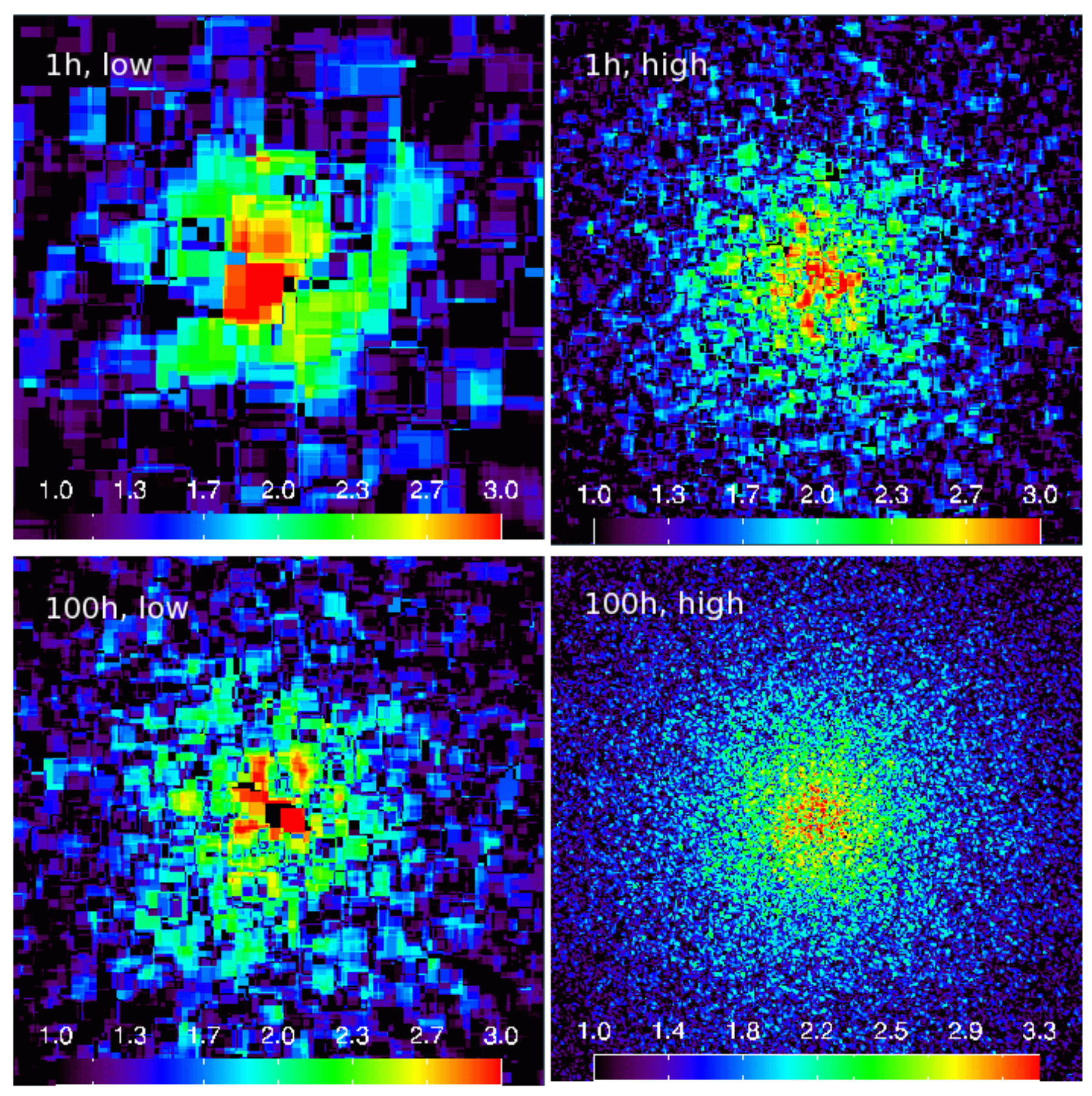} 
\caption{Simulated {\it SKA} Faraday rotation measure maps for a
turbulence-dominated (model B) cluster, in four observational
scenarios. Panel size and color scale are the same as in previous
figures.}\label{fig_kolmog}}
\end{figure*}
\begin{figure*}[t]
\center{
\includegraphics[width=0.75\textwidth]{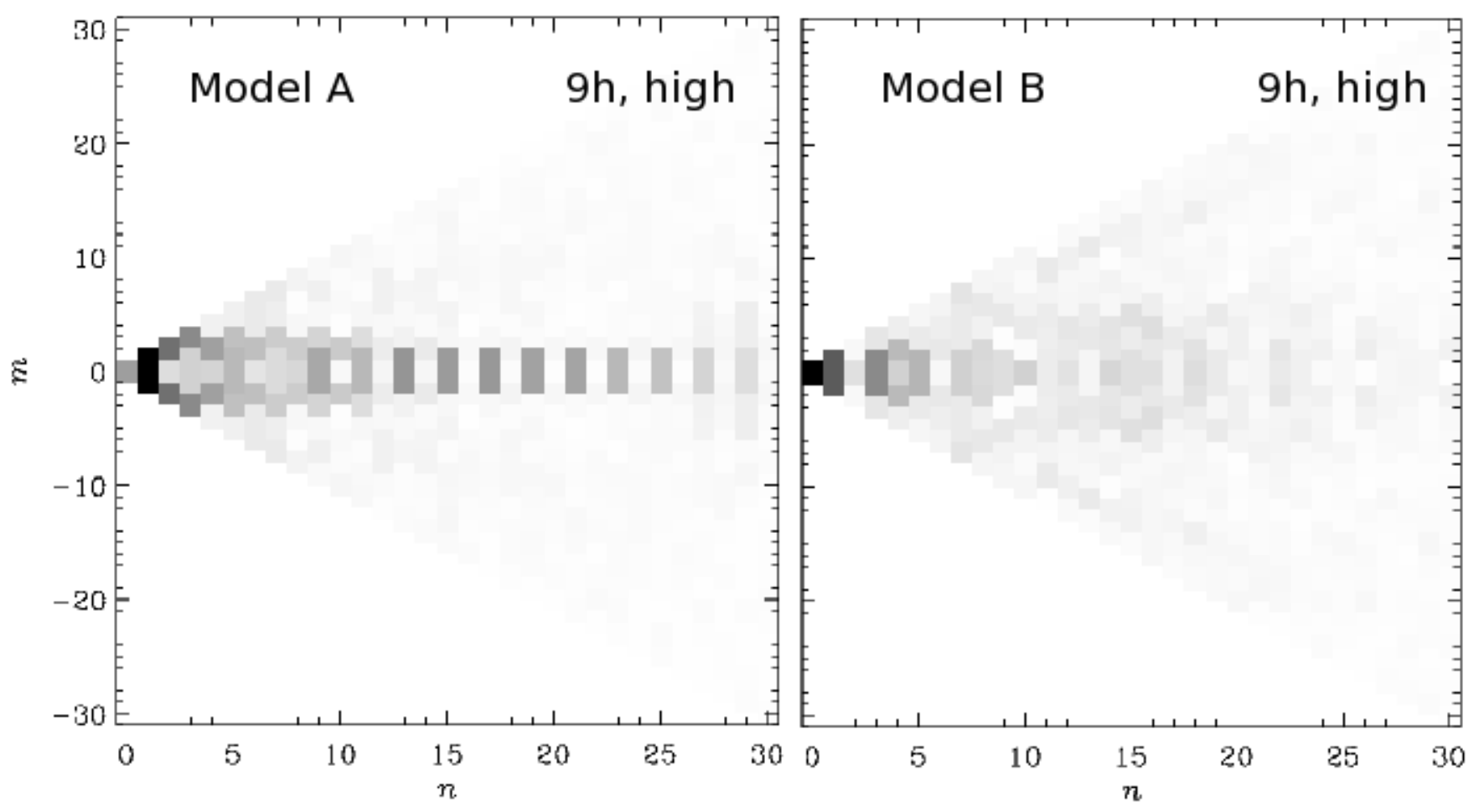} 
\caption{Polar shapelet decomposition of the RM images for model A
(left) and model B (right) in observational scenario {\it 9h, high}
with {\it EVLA}.  The amplitude of multipole coefficients is indicated
by the linear greyscale, with darker colors corresponding to higher
amplitudes. The (complex) shapelet coefficients also have phases,
which indicate the orientation of each multipole, but these are not
shown.}\label{fig_evla_decomp}}
\end{figure*}
\begin{figure*}[t]
\center{
\includegraphics[width=0.75\textwidth]{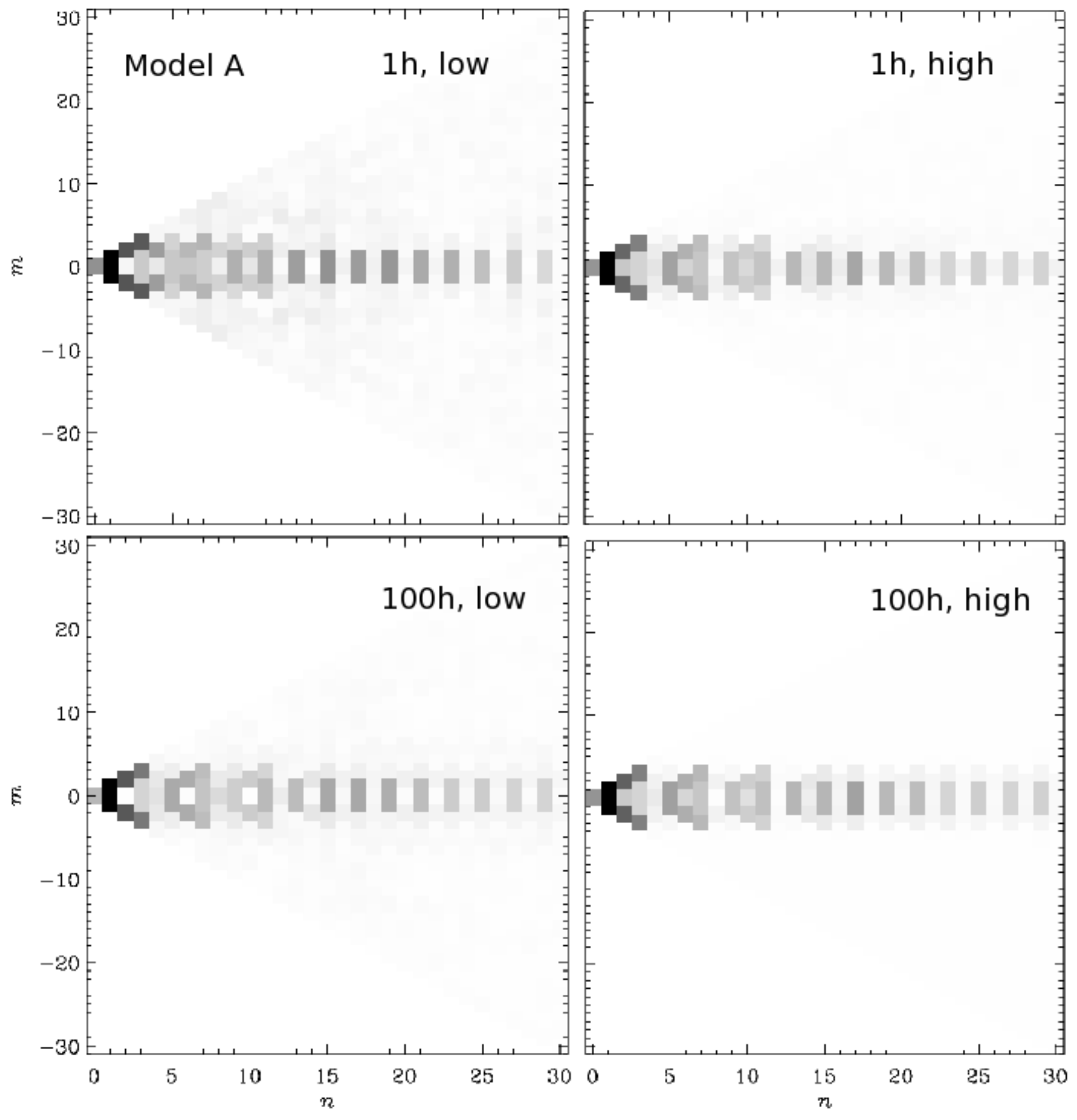} 
\caption{Polar shapelet decomposition of the RM images in model A
(instability-dominated cluster) for four different observational
scenarios with the {\it SKA}. Color scale is the same as in
Figure~\ref{fig_evla_decomp}.}\label{fig_hbi_decomp}}
\end{figure*}
\begin{figure*}[t]
\center{
\includegraphics[width=0.75\textwidth]{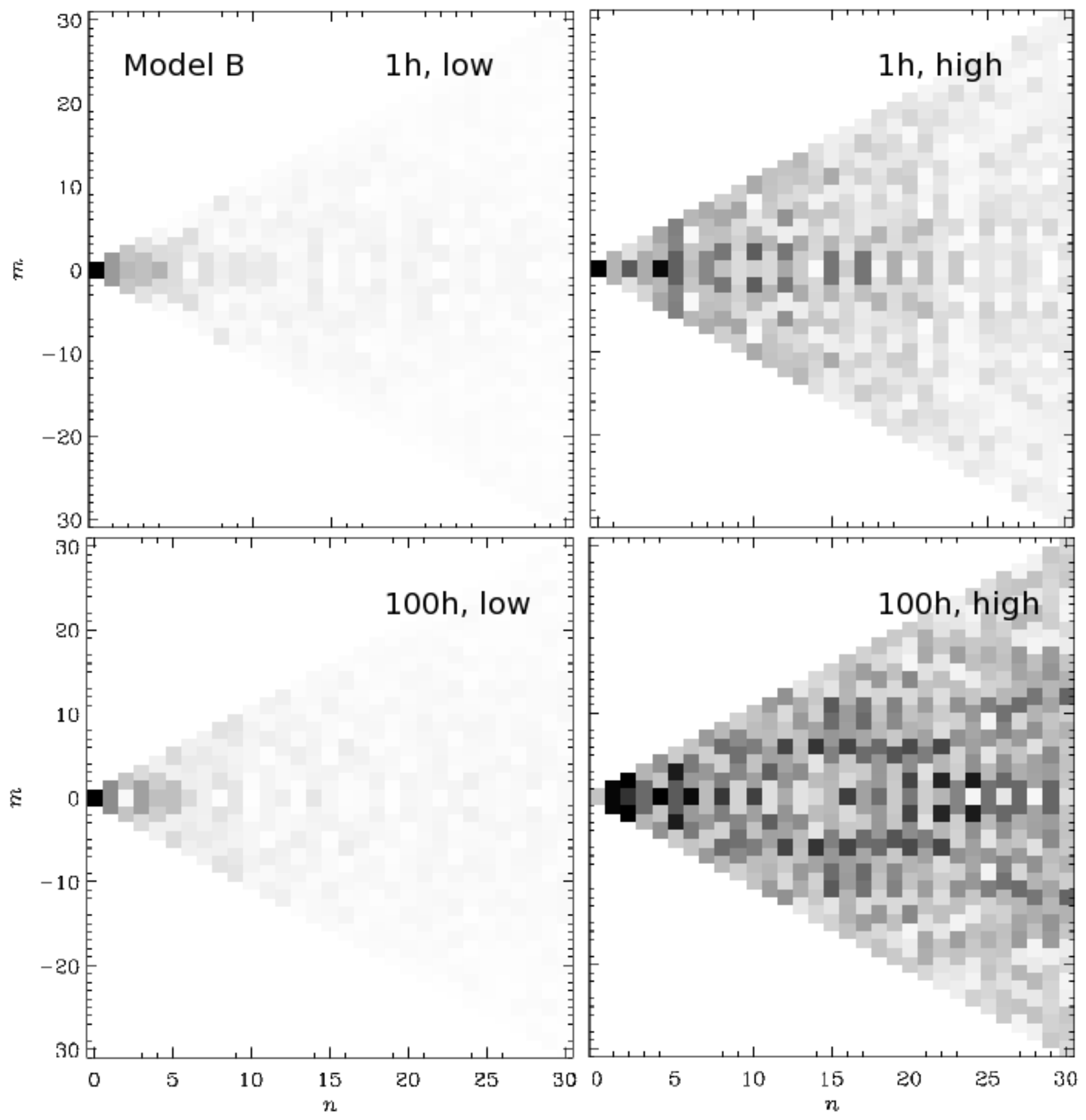} 
\caption{Polar shapelet decomposition of the RM images for model B
(turbulence-dominated cluster) for four different observational
scenarios with the {\it SKA}. Color scale is the same as in previous
figures.}\label{fig_kolmog_decomp}}
\end{figure*}

Given the {\it EVLA} and {\it SKA} sensitivity limits, we follow
approach in \citet{stepanov08} to estimate the number density of
polarized background sources at 1.4~GHz expected for our model
cluster. \citet{stepanov08} extrapolate the source counts at 1.4~GHz
from $P=0.5$~mJy to the limiting flux density of $P_{\rm min} =
0.05\,\mu Jy$ using a power-law relation between the cumulative number
of sources per square degree and polarized flux density,
$N_\Box(>P)\propto P^{-\gamma}$. The exponent $\gamma$ can take a
range of values between 0.7 and 1.1, which correspond to the low and
high bounds for the source counts at a given flux limit at 1.4~GHz. We
consider exposure times of 9h in case of {\it EVLA} and 1h and 100h
for {\it SKA}, both in the low and high source count scenarios, thus
obtaining a total of six different observational scenarios for both
models A and B. In Table~\ref{table} we list the total number of
background polarized sources ($N_s$) with polarized flux densities $>
5\sigma_{P}$ above the noise, within a $600\,{\rm kpc}\times 600\,{\rm
kpc}$ area centered on the cluster. Also shown is the mean separation
between the sources ($d_s$) at the redshift of the cluster.

Note that in the estimation of the number of background polarized
sources we simply extrapolated the number of sources from $0.5$~mJy
down to the microJy and nanoJy flux densities. The nature of the
dominant population of sources that {\it EVLA} and {\it SKA} will
detect at such low flux densities is uncertain, as they lie beneath
the surface of the deepest radio surveys that are currently
available. It has been proposed that microJy and nanoJy sky is
dominated by radio-quiet AGN and star-forming galaxies
\citep{jr04,wilman08} as well as faint ellipticals and dwarf galaxies
\citep{padovani10}. It is also worth pointing out a simplification in
our modeling which stems from an assumption that all background
polarized sources are point like, while in reality some low redshift
background sources will have a resolved extended structure. We
nevertheless expect these to comprise a small fraction, as a majority
of sources are expected to reside at high redshifts.

\section{Faraday rotation measure maps}\label{S_RMmaps}
\subsection{Construction and properties}\label{S_construction}

The polarization of electromagnetic waves traveling through the
cluster's magnetic field is rotated by an angle
$\Delta\chi=RM\,\lambda^2/(1+z)^2$, where $\lambda$ is the observing
wavelength and $z$ is the redshift of the Faraday screen.  We
calculate the effect of this Faraday rotation for our model cluster by
numerically integrating the expression RM$= 812\,{\rm rad\,m^{-2}}\int
n_e \, {\bf B}\cdot{\bf dl}$, where $n_e$ is the electron number
density in units of ${\rm cm^{-3}}$, ${\bf B}$ is the vector of
magnetic field in $\mu {\rm G}$, and {\bf l} is the vector demarking
the depth of the magnetic screen in kiloparsecs as measured along the
line of sight of an observer placed at infinity, at an arbitrary
orientation with respect to the cluster. The result is a two
dimensional, continuous map of RM intensity with $\sim 1\arcsec$
resolution shown in Figure~\ref{fig_mapsAB}. We convolve this map with
a map of point like, randomly distributed polarized background
sources, where the number of sources in the map is $N_s$. For visual
representation in Figures~\ref{fig_evla}, \ref{fig_hbi}, and
\ref{fig_kolmog} the convolved maps are smoothed with a box function,
where the size of each box is chosen so that it contains 5 polarized
background sources on average. This implies smoothing box of the size
$h_{\rm box} = 69$~kpc for the {\it EVLA} {\it 9h, high} scenario and
$h_{\rm box} = 60$, $15$, $26.4$, and $4.2$~kpc for the {\it SKA}
scenarios {\it 1h, low}, {\it 1h, high}, {\it 100h, low}, and {\it
100h, high}, respectively. We list the properties of the RM maps in
Table~\ref{table}.

Figure~\ref{fig_evla} shows maps for the models A (left panel) and B
(right panel) calculated for a 9h exposure with {\it EVLA} in the
scenario with an assumed high count of background polarized
sources. The RM structure in both appears marginally resolved. The
{\it 9h, low} scenario results in unresolved RM structure (the RM map
for this case is not shown but its properties are listed in
Table~\ref{table}, for comparison). This is not surprising given the
low number density of background sources within the considered area
with average spacing between the sources of $d_s\approx 75$~kpc,
demarking the effective size of the ``resolution element'' in this
scenario. Figures~\ref{fig_hbi} and \ref{fig_kolmog} show the RM maps
for models A and B, realized in each of the four observational
scenarios with the {\it SKA}. In both, the instability-dominated and
turbulence-dominated scenarios, the quality and spatial sampling of
the maps increase with the density of background polarized sources, to
the degree that the most optimistic observational scenario with the
{\it SKA}, {\it 100h, high}, almost exactly replicates the features of
the continuous RM maps shown in Figure~\ref{fig_mapsAB}. This is
implied by the hierarchy of characteristic scales, $\theta_{\rm SKA} <
d_s < \lambda_{\rm min}$, meaning that both the angular resolution of
the {\it SKA} and the density of the background polarized sources are
sufficient to fully capture the spectrum of the $\sim$kpc scale RM
variations in this model.

In all observational scenarios of model A it is possible to discern
two distinct regions, which are characterized by the HBI (core) and
MTI (outer region) instabilities. The dark swirl patterns in the
cooling core region, with low values of RM, arise where the magnetic
field changes direction along the line of sight. Because of the
multiple magnetic field reversals along the lines of sight, the RM
intensity ridges in the core do not correspond to the magnetic field
structure in a trivial way. Nevertheless, they carry an imprint of the
azimuthal distribution of the field lines, and if indeed present in
cooling core clusters, may be one of the characteristic features to
search for in RM surveys of clusters.  Outside of the core, where the
magnetic field geometry is radial with decreasing magnitude, the RM
intensity also decreases smoothly with radius. The diagonal feature
apparent in maps calculated for model A, in the left panel of
Figure~\ref{fig_evla} and all panels of Figure~\ref{fig_hbi}, is an
artifact of our model: the underlying symmetry in the radial component
of the magnetic field causes a cancellation in the RM along these
lines of sight resulting in low values of RM. The black specs occur in
places where the number of background sources per smoothing box falls
to zero. The lower cutoff value applied in maps ($10\,{\rm
rad\,m^{-2}}$) is comparable to the maximum error of RM reached at the
limiting flux density $P_{\rm min} = 5\sigma_P$ with the {\it EVLA}
and {\it SKA} \citep{stepanov08}.

In model B, shown in the right panel of Figure~\ref{fig_evla} and
Figure~\ref{fig_kolmog}, the RM patterns exhibit a noticeably
different, patchy distribution. The local maxima (minima) of the RM
distribution are again associated with regions where magnetic field
reversal results in an enhancement (a cancellation) along the line of
sight.  The main difference in Model B is that the cancellation effect
is more pronounced relative to model A, due to the tangled geometry of
the magnetic field, even though the assumed magnetic field strength is
comparable in both models (see \S~\ref{S_cluster_model}). This results
in the maximum RM value about an order of magnitude lower than in the
model A, where the magnetic field lines exhibit uniformity and
azimuthal structure on large scales.

\subsection{Discrimination between cluster models}\label{S_discrimination}

An image analysis technique well suited to characterizing the
different geometric patterns in the RM maps is {\it polar shapelets}
\citep{mr05}. This is based on the unique decomposition of localized
objects into a series of orthogonal basis functions that explicitly
separate modes with different rotational symmetries. Useful forms
exist in both 2-d and 3-d.  The former can generally be used for
accurate object photometry and astrometry \citep{kuijken06}, as well
as morphological classification of the images of galaxies
\citep{km05,massey07}, magnetograms of sunspots \citep{young05}, and
the response of the human visual cortex \citep{victor09}. Its
convenient mathematical properties and intuitive interpretation also
make it a particularly effective morphology estimator for clusters'
magnetic fields.

We fit each image $I({\mathbf x})$ in Figures~\ref{fig_evla},
\ref{fig_hbi} and \ref{fig_kolmog} as a weighted sum of shapelet basis
functions $\chi_{n,m}(x,y)$ such that
\begin{equation} \label{eqn:sseriesp}
I(x,y) = \sum_{n=0}^{30} \sum_{m=-n}^{n} a_{n,m}\,\chi_{n,m}(x,y) ,
\end{equation}
where the (complex) coefficients $a_{n,m}$ describe the power in modes
with $n$ radial oscillations and $m$-fold rotational symmetry, similar
to a localized Fourier transform. Higher $n$-orders also capture
structure at increasing distances from the cluster core. In general,
the sum over $n$ can extend to infinity, although in practice we have
arbitrarily truncated it to 30. The sum over $m$ need include only
every other term, because the intervening basis functions are
explicitly zero. We focus our analysis on the inner 600~kpc region of
the modeled cluster, which encloses the most interesting RM patterns,
and show the magnitudes of the derived shapelet coefficients $a_{n,m}$
in Figures~\ref{fig_evla_decomp}, \ref{fig_hbi_decomp},
\ref{fig_kolmog_decomp}.

The analysis of the maps for different observational scenarios within
model A (or model B) give qualitatively similar results, but with
varying ratios of signal to noise. The key idea is that the shapelet
decomposition efficiently isolates modes of rotational symmetry
present in the RM maps, even in the presence of significant
observational noise. Indeed, the distribution of power in the two
models is strikingly different. The shapelet decomposition of
scenarios in model A shows almost all power within the range
$m=\{-3,3\}$, while it fans out uniformly over $m$ in turbulent model
B.  The radially symmetric ($m=0$) modes describe the average, uniform
level of the RM intensity.  The restriction of model A's deviations
from dipole ($|m|=1$), quadrupole ($|m|=2$) and sextupole ($|m|=3$)
modes with low orders of rotational symmetry can be understood in
light of the equivalent $\theta$ and $\phi$ terms in
equations~\eqref{eq_Btheta} and \eqref{eq_Bphi}. For each model,
Table~\ref{table} lists the fraction of power in shapelet coefficients
with $|m|=0$ to 3, as estimators similar to the shapelet asymmetry
estimator \citep[equation 61 of][]{massey07}.

In all the observational scenarios that we have considered, the
magnetic field geometry of model A is distinguished from that of model
B as a large fraction of RM power in modes $|m|\leq 3$ and a
decreasing power in higher-$|m|$ modes. In scenarios with a low
density of background polarized sources, namely {\it 9h, high} (left
panel of Figure~\ref{fig_evla_decomp}) and {\it 1h, low} (top left
panel of Figure~\ref{fig_hbi_decomp}), the observed RM measure map has
insufficient resolution to contain all the $|m|=3$ modes, so the
cutoff is present but less distinct. That the two {\it EVLA} and {\it
SKA} observational scenarios are indeed similar is indicated by the
comparable polarized source densities and similar distributions of the
RM intensity (Table~\ref{table}). Most excitingly, even in observing
scenarios with noisier data, which may be achieved in longer exposures
with {\it EVLA} and relatively short exposures with {\it SKA}, the
large-scale magnetic field patterns are efficiently captured by polar
shapelets. In the remaining observational scenarios for the {\it SKA},
characterized by the higher density of background polarized sources
({\it 1h, high}, {\it 100h, low} and {\it 100h, high}), the
distribution of power remains very robust, as illustrated by the polar
shapelets decomposition.

In model B, the power is more uniformly distributed among the
azimuthal modes ($m$-modes) while the distribution across the radial
modes ($n$-modes) changes with the number density of background
polarized sources. In observational scenarios with the lower density
of sources ({\it 9h, high}, {\it 1h, low} and {\it 100h, low}), a
significant fraction of the power is in lower $n$-modes and thus, the
RM structure is captured on larger scales and not captured on small
scales. With the increasing density of polarized sources some of the
power shifts uniformly towards higher $m$ and $n$-modes (see {\it 1h,
high} scenario), indicating that finer RM structures are beginning to
be resolved in this scenario. 

A practical implementation of the shapelet decomposition also requires
advance estimates of a cluster's center and size, so that the basis
functions can be constructed at a given location.  For this analysis,
we iteratively optimized the size of the basis functions to minimize
residuals between the simulated RM map and its shapelet model.  The
scale factor of the basis functions represents the size of the
Gaussian used in image deconstruction. In combination with the maximum
(truncation) value of $n$ in the model, it determines its resolution
in such way that smaller values of the scale factor allow higher
resolution models. The same scale size also sets the maximum spatial
extent of the model, and there is a balance between the ability to
model the large scale features and the high frequency detail, a choice
that is optimized via $\chi^2$ minimization. The used sizes of the
basis functions thus naturally arise to be different in the various
observational scenarios and between models A and B, depending on the
detail of the RM patterns.  This approach allows to constrain the
characteristic scale of the smallest (resolved) RM patterns using this
iterative procedure, however, our results are robust to setting a
fixed physical scale size to the reconstruction. Also for the purposes
of this analysis, we assumed that the center of a cluster would be
known {\it a priori}.  It would be possible to determine the best-fit
center via iteration on the RM image itself, but it will likely be
known to better accuracy in practice from independent (e.g.\ optical
or X-ray) observations.  Spurious offsets in the center primarily
shift power in the dipole $|m|=1$ shapelet modes.  Indeed, small-scale
structure near the cluster core in this realization, amplified by the
high electron density, is responsible for the large $a_{1,\pm 1}$ and
$a_{3,\pm 3}$ coefficients, which correspond to basis functions with a
small spatial extent.  If the center were determined from this data
alone, this structure would pull the center around, shifting some
power between adjacent shapelet coefficients \citep{massey07}.

We emphasize that the exact distribution of power in decomposed RM
images is model-dependent and that neither of our two models should be
regarded as a strict prediction of future observations. More
generally, our analysis demonstrates that physical mechanisms that
qualitatively adjust the magnetic field distribution in clusters can
be easily distinguished using observed RM maps, with plausible
exposure times.

\section{Depolarization and diffuse foreground emission}\label{S_depolarization}
 
While the sensitivity of the instrument and (consequently) the density
of observable polarized background sources strongly affect the
precision of Faraday rotation measure observations, the ability to tie
the observed properties of the RM maps to the underlying magnetic
field also depends on the effect of depolarization. Depolarization is
a reduction of the observed degree of polarization which may arise
within the cluster itself due to its spatial extent (internal
depolarization) or due to limitations in instrument capabilities
(beamwidth and bandwidth depolarization). We now disucss the
importance of beamwidth and bandwidth depolarization in the context of
our models.

{\it Beamwidth depolarization} arises when the minimum magnetic field
coherence length is smaller than the beam size of the radio
instrument, and cancellation of the RM occurs within the beam. Given
that the assumed beamwidth in our calculation is $1\arcsec = 353\,$pc
and the magnetic field coherence lengths are much larger ($\geq
7.5$~kpc), no beamwidth depolarization is expected to occur in these
scenarios.  A similar effect that can lead to a small loss of
information is smoothing that we apply to all of our RM images. In our
simulated observations, the smoothing length varies between
$\sim4$~kpc and $\sim70$~kpc, depending upon the density of background
sources in a particular observational scenario, and this determines
the size of the smallest RM patterns that can be inferred with
confidence from a smoothed image. However, this does not represent a
fundamental limitation, since more sophisticated smoothing methods and
scales could be adopted on real data.

{\it Bandwidth depolarization} is the cancellation and averaging of
the Faraday rotation measure that arises in polarimeters operating in
fixed, wide frequency bands. Bandwidth depolarization will be largely
eliminated by the {\it EVLA} and {\it SKA}, by acquiring RM
measurements via the rotation measure synthesis method. RM-synthesis
is based on multichannel spectro-polarimetry to enable the detection
of weak, polarized emission \citep{bb05b}. Most importantly, this also
allows the simultaneous observation of a range of different RM values,
and the separation of RM components from distinct regions (such as
foreground and background structures) along the line of sight.
Indeed, the RM signal due to Faraday rotation of polarized light from
real background sources is complicated by diffuse, intrinsic polarized
emission from both the ICM plasma itself, and our own Galaxy.

{\it Diffuse, polarized foreground emission} from cooling core
clusters can be attributed to mini-halos, steep spectrum radio sources
associated with the ICM around a powerful central radio galaxy. The
prototypical example is a $\sim450$~kpc mini-halo at the center of
Perseus. \citet{bren11} find that the diffuse emission observed at
350~MHz in the direction of Perseus seems not to be related to the
mini-halo but rather to the foreground emission from the Milky
Way. Our Galaxy interferes with measurements by Faraday-rotating any
extragalactic polarized signal and by adding its own polarized
emission. RM of Galactic origin is typically $\sim10\,{\rm rad\,
m^{-2}}$ but can be as high as $300\,{\rm rad\, m^{-2}}$ for objects
close to the Galactic plane \citep{sn81}. In the direction of the
Perseus cluster for example, \citet{bren11} measure a relatively low
RM contribution (compared to our modeled values) in the range -50 to
$+100\,{\rm rad\,m^{-2}}$, which can in principle be disentangled from
the RM map of the cluster using the RM-synthesis technique and is not
expected to produce an RM signal competing with that from the cluster.
Thus, we conclude that the RM features modeled in this work are not
expected to be significantly affected by depolarization and should
dominate in magnitude over any component of RM contributed by the
sources of diffuse polarized emission. The primary factor that
determines the efficacy of the RM maps in probing the magnetic field
structure is the sensitivity of the spectro-polarimetric measurements.

\section{Discussion and conclusions}\label{S_discussion}

Even if MTI and HBI instabilities operate uninhibited in real
clusters, the radial-azimuthal field geometry is likely to be
perturbed by intermittent phases of AGN activity and mergers. As these
perturbations are attenuated over time, they can source a new cycle of
MHD instabilities. The relics of such events, including shocks,
bubbles, and ridges of magnetic field lines swept by intracluster
galaxies, may also be recognizable in high quality RM maps --
testifying to the past evolution of a cluster. But with a
characteristic time scale for the saturation of MHD instabilities of
only a few billion years, the magnetic field lines can be driven to
recover their orientation relatively quickly. As a consequence, a
fraction of clusters may exhibit preferential magnetic field
geometries, despite episodic disruptions.

Circumstantial evidence in support of this hypothesis may already
exist in several forms.  Recent detailed observations of the Virgo
cluster revealed an isothermal region with remarkable azimuthal
symmetry cocooned in the cool core of the cluster between the active
radio-lobes of M87 \citep{million10}.  The metallicity of the
isothermal gas in the same region is non-uniform and clumpy. It is
reasonable to expect that the high metallicity parcels of gas were
uplifted from the low entropy, cool core region of the cluster, where
the density of stars is highest. That the parcels now have the entropy
and temperature of the surrounding ambient plasma, but retain clumpy
metallicity distribution, suggests that turbulent mixing was not
efficient here and that the gas was instead heated by conduction.  The
presence and shape of the isothermal front indicate that the magnetic
field geometry in this region may be {\it predominantly azimuthal}.
Furthermore, the absence of efficient turbulence indicates that HBI
may operate unhindered in this region and that the instability may be
more robust than suggested by some recent theoretical works
\citep{ro09, pqs10}. However, at least one more physical phenomenon
may produce a similar magnetic field topology: \citet{ro09} find that
weak turbulent motions lead to trapped $g$-modes and result in gas
motions that are preferentially tangential. They suggest that magnetic
fields in clusters experiencing such $g$-modes can in principle become
tangential even in the absence of thermal conduction and the
HBI. While some thermal conduction seems to be implied in the case of
the Virgo cluster, the concurrent presence of $g$-modes cannot be
eliminated.

\citet{juett10} draw attention to what may be more circumstantial
evidence for MHD instabilities.  As much as $20\%$ of the sample of 70
clusters presented by \citet{snowden08} have puzzling temperature
profiles that appear to be quasi-isothermal at $\sim$Mpc
radii\footnote{The classification of the sample of quasi-isothermal
clusters according to the thermal state of their cores has not been
reported thus far.}. The clusters show no signs of interaction, which
could possibly have accounted for their temperature distribution, and
they appear dynamically relaxed. Moreover, it is unclear whether
models based on theoretical studies of cluster properties could
account for this unusual class of objects \citep{nagai07}. If the
class is shown to be unexpected, it may point to some missing physics
in cosmological simulations. A possible explanation for the
temperature structure of these objects is that thermal conduction
operates very efficiently in their outer regions, an effect that
arises as a natural consequence of the MTI instability. Radial
configuration of magnetic field lines in outer regions of such
clusters could only persist if they are isolated and unperturbed for
sufficiently long periods of time, consistent with the observed
properties of the quasi-isothermal clusters. Another line of evidence
comes from the study of magnetic field structure around galaxies in
the Virgo cluster. \citet{pd09} find that in Virgo, which seems to be
in transition to a cool core, the global magnetic field has a
predominantly radial orientation at large radii, which again suggests
the operation of MTI in its ICM.

In this study, we have evaluated the effects of two different physical
mechanisms on the Faraday rotation measure of a magnetized cooling
core cluster, in the context of the planned capabilities of the {\it
EVLA} and {\it SKA} radio observatories. We compare a theoretical
scenario in which conduction-driven MHD instabilities dominate the
dynamics of the ICM, to a scenario in which magnetic field topology is
defined by turbulent motions. We employ the polar shapelets image
analysis method to efficiently detect patterns in the RM image with
specific rotational symmetries, and thus classify their
morphologies. Within the bounds of our simple models we find that the
two mechanisms can produce strikingly different RM patterns and that
future spectro-polarimetric measurements will have sufficient
sensitivity to discriminate between them that can be achieved in
longer exposures with {\it EVLA} and relatively short exposures with
{\it SKA}. We propose that the effect of the HBI and MTI instabilities
be sought for in dynamically relaxed cooling core clusters, and
especially in the subclass of clusters with quasi-isothermal
temperature profiles at large radii.  More generally, it should be
possible to discern physical mechanisms that result in qualitatively
different magnetic field topologies from observed Faraday rotation
measure maps, without a priori knowledge about the nature of the
processes. Such observations will enable detailed investigations into
the behavior of MHD instabilities and other associated physical
phenomena, which are of far reaching importance to a number of
fundamental questions related to energy transport in clusters.

\acknowledgments

We thank the anonymous referee for thoughtful comments which helped to
significantly improve this manuscript. T.B. would like to thank Tracy
Clarke and Richard Mushotzky for stimulating discussions and useful
suggestions. Support for T.B. was provided by the National Aeronautics
and Space Administration through Einstein Postdoctoral Fellowship
Award Number PF9-00061 issued by the Chandra X-ray Observatory Center,
which is operated by the Smithsonian Astrophysical Observatory for and
on behalf of the National Aeronautics Space Administration under
contract NAS8-03060. T.B. and C.S.R. acknowledge support from the NSF
under grant AST-0908212.  R.M.\ is supported by STFC Advanced
Fellowship PP/E006450/1 and FP7 grant MIRG-CT-208994.


\end{document}